\documentclass[11pt]{article}

\usepackage{amsmath,amssymb,theorem,enumerate,fullpageK,epsfig}

\theoremstyle{plain}
\newtheorem{proposition}{Proposition}
\newtheorem{theorem}{Theorem}
\newtheorem{lemma}{Lemma}
\newtheorem{corollary}{Corollary}

{{\theorembodyfont{\rmfamily}
\newtheorem{definition}{Definition}
\newtheorem{remark}{Remark}}

\newcommand{\byeproof}{\hfill $\blacksquare$}

\newcommand{\beginE}{\begin{equation}}
\newcommand{\closeE}{\end{equation}}

\newcommand{\bbR}{{\mathbb{R}}}

\newcommand{\pd}{\partial}

\newcommand{\inner}[2]{\langle #1, \; #2 \rangle}
\newcommand{\eps}{\varepsilon}

\renewcommand{\a}{\alpha}
\renewcommand{\b}{\beta}

\newcommand{\myand}{\quad \mbox{and} \quad}
\newcommand{\kong}{\varnothing}
\newcommand{\lead}{{\mathcal{L}}}

\begin{document}

\setlength{\baselineskip}{14pt}

\title{Cucker-Smale Flocking under Hierarchical Leadership}

\author{Jackie (Jianhong) Shen
        \thanks{This cover page is only for the purpose of submission. Work has been
        partially supported by (USA) NSF under grant No. DMS-0604510 (2006-2009). } \\
        School of Mathematics, University of Minnesota, Minneapolis, MN 55455, USA \\
        {\tiny \bf Email: jhshen@math.umn.edu; \quad Tel. (612) 625-3570}
        }
\date{}

\maketitle

\begin{abstract}
A mathematical theory on flocking serves the foundation for several ubiquitous
multi-agent phenomena in biology, ecology, sensor networks, economy, as well as social
behavior like language emergence and evolution. Directly inspired by the recent
fundamental works of Cucker and Smale on the construction and analysis of a generic
flocking model, we study the emergent behavior of Cucker-Smale flocking under {\em
hierarchical leadership}. The rates of convergence towards asymptotically coherent group
patterns in different scenarios are established.

The consistent convergence towards coherent patterns may well reveal the advantages and
necessities of having leaders and leadership in a complex (biological, technological,
economic, or social) system with sufficient intelligence. \vskip 6pt

\noindent {\bf Key words.} Bio-flocking, leaders, leadership, Cucker-Smale model, dynamic
graphs, graph Laplacian, Fiedler number, convergence, perturbation, free will. \vskip 6pt


\end{abstract}

\tableofcontents

\newpage

\section{Introduction and Motivations}

\subsection{General Background on Flocking}

Flocking, a universal phenomenon of multi-agent interactions, has gained increasing
interest from various research communities in biology, ecology, robotics and control
theory, sensor networks, as well as sociology and economics.
 \begin{enumerate}[(i)]
 \item ({\bf Biology and Ecology}) The emergent behavior of bird flocks, fish schools, wolf packs,
 elephant herds, or bacteria aggregations, for example,
 has long been a major research topic in population and behavioral
 biology and ecology~\cite{BertozziPRL06,CouzinLeader05,CuckerSmaleA,CuckerSmaleB,FlierlGrunbaumLevin99,LiuPassino04,TopazBertozziSIAM04,TopazBertozziLewis05}.

 \item ({\bf Robotics and Control}) The coordination and cooperation among multiple mobile agents
 (robots or sensors)
 have been playing central roles in sensor networking, with broad applications in
 military, environmental control, or various field tasks~\cite{JadbabaieLinMorse03,Tsitsiklis86}.

 \item ({\bf Economy and Languages}) Emergent economic behavior, such as a common belief in a
 price system in a complex market environment, is also intrinsically connected to
 flocking.  The emergence of a common language in primitive societies is yet another
 example of a coherent collective behavior emerging within a complex
 system~\cite{CuckerSmaleB,CuckerSmaleZhouL04}.
 \end{enumerate}

The present work can largely be categorized into the biology realm, and has been directly
inspired by the recent mathematical works of Cucker and
Smale~\cite{CuckerSmaleA,CuckerSmaleB}, as the title suggests. Mathematical abstraction
and rigorous analysis are more focused herein than actual  biological or physical
realizability or feasibility. As in physics, the study of idealized models can often shed
light on various observed patterns in the real world, {\em if\;} such models can indeed
catch the very essence.

In biology and physics, the main goal of flocking study is to be able to interpret,
model, analyze, predict and simulate various flocking or multi-agent aggregating
behavior. Most existing works have been focusing on modeling and
simulation~\cite{LevineRappel00,Vicsek95}. See, for example, the several important models
investigated by Flierl et al.~\cite{FlierlGrunbaumLevin99} (and their stochastic
formulation). The more recent paper of Parrish et al.~\cite{ParrishViscidoG02} also
provides a comprehensive comparison among some major existing models and their governing
variables (in the context of fish schooling). Quantitative analysis (as
in~\cite{CuckerSmaleA,CuckerSmaleB,JadbabaieLinMorse03}) on the asymptotic rates of
emergence and convergence, on the other hand, has been relatively rare.

Mathematical efforts are gradually gaining strength in this multidisciplinary area. In
the continuum limit, for example, there have been several recent efforts made by
Bertozzi's group~\cite{TopazBertozziSIAM04,TopazBertozziLewis05}, in which global
swarming (i.e., with densely populated agents) patterns are modeled and analyzed via
suitable spatiotemporal differential equations. Discrete-to-continuum limits of
interacting particle systems have also been investigated by the same
group~\cite{BertozziPhysD06,BertozziPRL06} recently. Consistent and generic mathematical
analysis has been very much in an early stage for many biological aggregation phenomena.
In the current paper, following the recent remarkable works of Cucker and
Smale~\cite{CuckerSmaleA,CuckerSmaleB} on flocking analysis, we attempt to make further
extension along the same line.

\subsection{Cucker-Smale Flocking Model}

Given a flock of $k$ agents (birds, fish, wolves, etc) labeled by $i=1,2, \dots, k$, the
Cucker-Smale flocking model is  specified by the {\em nonlinear} autonomous dynamic
system:
 \beginE \label{E1:CSmodel}
 \begin{cases}
 \dot x_i(t) = v_i, \\
 \dot v_i(t) = \sum_{j \in \lead(i)} a_{ij}(x) (v_j -v_i), \qquad i=1:k, t>0,
 \end{cases}
 \closeE
where $x_i(t)$ and $v_i(t)$ are 3D (3 dimensional, which is non-essential) position and
velocity vectors at time $t$, $x=(x_1, \dots, x_k) \in (\bbR^3)^k$, and $\lead(i)
\subseteq \{1,\cdots k\}$ denotes the subgroup of agents that directly influence agent
$i$. Furthermore, the {\em connectivity coefficients} $a_{ij}(x)$ are in the form of
 \[ a_{ij}(x)= w(|x_i -x_j|^2), \qquad \mbox{for some nonnegative weight profile $w(y)$.} \]
In the current paper, by {\em Cucker-Smale flocking model}, we require as
in~\cite{CuckerSmaleA,CuckerSmaleB} that the interaction weight function $w(y)$ takes the
form of:
 \beginE \label{E1:wchoice}
    w(y) = \frac {H}{(1 + y)^\b}, \qquad \mbox{or} \qquad
    w(y) \ge \frac {H}{(1 + y)^\b},
 \closeE
where $H$ and $\b$ are two positive system parameters. One shall see that the two ($=$
vs. $\ge$) make no difference for the analysis hereafter as long as $w(y)$ is bounded and
sufficiently smooth (also see~\cite{CuckerSmaleA}). We also must point out that this
model has been put in a more general and abstract setting in the subsequent work of
Cucker and Smale~\cite{CuckerSmaleB}.

The look of the system~(\ref{E1:CSmodel}) is not entirely new. For example, the 2D model
studied by Vicsek et al.~\cite{Vicsek95} is very similar in which $v_i$'s share the same
magnitude (or speed) while their heading directions $\theta_i$'s satisfy a similar set of
equations.

It is the particular choice of the connectivity coefficients in~(\ref{E1:wchoice}) that
has made the Cucker-Smale model mathematically more attractive. Vicsek et al.'s
 model (in discrete time)~\cite{Vicsek95} can be considered as taking the following cut-off
weight function:
 \[
 w(y) = w_r(y) = 1_{y \le r^2}(y), \qquad \lead(i) \equiv \{1,\dots,k\}, \quad \forall
 i.
 \]
That is, two distinct agents $x_i$ and $x_j$ interact if and only if they are within a
distance of $r>0$, which is assigned a priori and fixed throughout. The lack of
long-range interactions has made the model very difficult to analyze. For example, the
remarkable efforts of Jadbabaie et al.~\cite{JadbabaieLinMorse03} on emergence analysis
avoided the actual dynamic dependence of $a_{ij}$ on the configuration $x$, but instead,
they focused on an altered setting that involves switching controls.

The main results of Cucker and Smale~\cite{CuckerSmaleA} can be summarized as follows:
when $\beta < 1/2$, the flock converge to some translating rigid structure (moving at a
constant velocity) {\em unconditionally}, i.e., regardless the initial configuration; and
when $\beta \ge 1/2$, the initial velocities and positions have to satisfy certain
compatible conditions so that the entire flock can converge asymptotically.

In summary, in the modeling and analysis of Cucker and
Smale~\cite{CuckerSmaleA,CuckerSmaleB}, not only are the conditions for pattern emergence
easily verifiable (i.e., by checking the initial conditions), but the role of long-range
interaction is also clearly quantified. A smaller $\beta$ signifies more intense
long-range interactions among agents while a bigger $\beta$ leads to much weaker ones. It
has been shown that the critical exponent $\beta_c=1/2$ is sharp and necessary.
Previously, the connection between global pattern emergence and individual action rules
has often only been observed experimentally or addressed empirically (Vicsek et
al.~\cite{Vicsek95}, for example, experimentally observed phase transition induced by
population density $\rho$ and random fluctuation $\eta$. A higher density corresponds to
more interaction among agents, or loosely, smaller $\beta$ in the Cucker-Smale model.)

\subsection{Motivations and Main Results of Current Work}

In the current work, we investigate the emergent behavior of Cucker-Smale flocking under
hierarchical leadership (HL), which will be defined in detail in the next section.

{\em Roughly, an HL flock is one whose members can be ordered in such a way that
lower-rank agents are led and only led by some agents of higher ranks}.  As explained in
more details in Section 2, for HL flocks, it is often either nontrivial or impossible to
define a ``fixed" inner product so that the Fiedler number of the associated (graph)
Laplacian can be exploited, which is the key to the original work of Cucker and
Smale~\cite{CuckerSmaleA} and its subsequent generalization~\cite{CuckerSmaleB}. The
current work thus takes a somewhat different approach in order to fully benefit from the
characteristic structures of HL.

As far as applications are concerned, there are two types of HL: passive and active ones.
 \begin{enumerate}[(A)]
 \item ({\bf Passive/Transient Leadership})
  \begin{enumerate}[{(A.}1{)}]
  \item ({\bf Disturbed Bird Flocks}) In nature, certain types of leadership emerge in a transient and
 dynamic fashion and is often prompted by a specific environment. For a disturbed bird
 flock at rest, for example, the bird that first senses the approach of an unexpected
 pedestrian or predator often takes flight first, warns others, and first gains full
 speed, and consequently flies ahead of the entire flock and serves as a virtual
 leader.

  \item ({\bf Driving in a Traffic}) During rush hours, each individual driver mainly
  manoeuvres according to the moving patterns of several cars right ahead in the visual field.
  Thus a chain of leadership naturally arises and extends linearly along the
  traffic. The leadership here is also prompted by the environment rather than being
  intrinsic among the stranger drivers.
  \end{enumerate}

 \item ({\bf Active/Intrinsic Leadership})
  \begin{enumerate}[{(B.}1{)}]
   \item ({\bf Governmental/Miltiary Hierarchies}) Such hierarchical leadership
   is inherent in various social groups or structures, and often leads to more efficient
   management. Examples include, the chain of President-Governor-Mayor in
   the governmental system, and the chain of command from the Commander in Chief all the
   way down to a soldier.

   \item ({\bf Social Animals}) For some social animals such as monkeys, wolves,
   or elephants~\cite{CouzinLeader05},  the
   group or social status of each member is clearly recognized by others and stably
   maintained, and guides the action of each individual in the hierarchies. (See also
   the recent work of Couzin et al.~\cite{CouzinLeader05} for non-hierarchical but
   ``effective" leadership.)
  \end{enumerate}
 \end{enumerate}

Our main results are the three theorems summarized below. All HL flocks are assumed to
have Cucker-Smale connectivity introduced in the preceding subsection.

\begin{enumerate}[(i)]
\item ({\bf Section 3}) For an HL $(k+1)$-flock marching at a sufficiently small
{\em discrete} time step $h$, under the similar classification scheme according to $\beta
<\b_c$, $=\b_c$, or $>\b_c$, as in Cucker and Smale~\cite{CuckerSmaleA,CuckerSmaleB}, the
velocities of the flock converge at a rate of $O(\rho_h^n n^{k-1})$, where the factor
$\rho_h \in (0, 1)$ only depends on $h$, system parameters, and the initial configuration
of the flock. The critical exponent is given by $\b_c = 1/(2k)$, instead of $\b_c =1/2$
in the original work of Cucker and Smale~\cite{CuckerSmaleA}. (For a $2$-flock (with
$k=1$) they are the same. For $k>1$, the $\b_c$ herein could be over restrictive and due
to the deficiency of the particular methodology adopted.)

\item ({\bf Section 4}) For an HL flock under continuous-time dynamics, when
$\b<1/2$, there exists some $B>0$, such that the velocities of the flock converge at an
exponential rate of $O(e^{-Bt})$. The constant $B$ only depends on the system parameters
and the initial configuration of the flock. (From the simple calculation on an HL
2-flock, $\b_c=1/2$ is sharp in order to achieve {\em unconditional} convergence.)

\item ({\bf Section 5}) For an HL $(k+1)$-flock $[0, 1, \dots, k]$ of which the overall
leader agent 0 takes a free-will acceleration $\dot v_0 =f(t)$ (thus the system is no
longer autonomous), as long as the overall leader behaves moderately so that
$f(t)=O((1+t)^{-\mu})$ for some $\mu > k$, the velocities of the flock will still
converge at a rate of $O( (1+t)^{-(\mu - k)})$ when $\b < 1/2$. (By (ii) where $f\equiv
0$, $\b_c=1/2$ is again sharp for unconditional convergence.)
\end{enumerate}

We also mention that Jadbabaie et al.~\cite{JadbabaieLinMorse03} also studied (under
discrete time and working with Vicsek et al.'s orientation model~\cite{Vicsek95}) the
effect of a {\em single} leader moving at a {\em fixed constant} velocity. As mentioned
above, due to the difficulty in dealing with configuration-dependent dynamics, the
authors switched to the study of an altered {\em control} problem (under the assumption
of intermittent joint connectivity).

In addition to the three main sections mentioned above, definitions and further detailed
background will be introduced in Section 2. The conclusion is drawn in Section 6.

\section{HL Flocks, and Definability of Compatible Inner Products}

\subsection{Flocks under Hierarchical Leadership (HL Flocks)}

\begin{definition}[An HL Flock]
A $(k+1)$-flock is said to be under {\em hierarchical leadership}, if the agents (birds,
fish, wolves, etc.) can be labeled as $[0, 1, \dots, k]$, such that
 \begin{enumerate}[(i)]
 \item $a_{ij}=a_{\mbox{agent $i$ led by $j$}} \; \neq 0$ implies that $j < i$; and
 \item if we define the {\em leader set} of each agent $i$ by
        \[ \lead(i)=\{ j \mid a_{ij} > 0 \}, \]
        then for any $i> 0$, $\lead(i) \neq \kong$ (non-empty).
 \end{enumerate}
If so, the flock is called \underline{\em an HL flock}.
\end{definition}

\begin{figure}[ht]
\centering{
 \epsfig{file=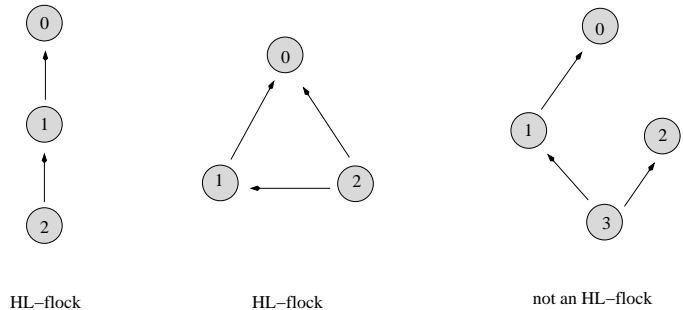, height=1.6in}
 \caption{Two examples of HL flocks and one example of a non-HL flock. The arrow
          $i \to j$ means that agent $i$ is led by agent $j$, or equivalently,
          $a_{ij} > 0$. Visually, it means that $i$ looks up to $j$.} \label{F2:HL}
}
\end{figure}

Notice that the second condition requires that, except for agent 0, all the others must
be subject to some leadership. On the other hand, the first condition implies that
$\lead(0)=\kong$. Thus agent 0 is the overall leader ({\em direct or indirect}) for the
entire flock. Figure~\ref{F2:HL} depicts the connectivity structure of two HL flocks and
one non-HL flock.

\begin{proposition}[Connectivity Matrix of an HL Flock]
A $(k+1)$-flock is an HL flock if and only if after some ordered labeling $[0, 1, \dots,
k]$, the connectivity matrix $K=(a_{ij})_{0\le i,j \le k}$ is lower triangular, and for
any row $i>0$, there exists at least one positive off-diagonal element $a_{ij}$.
\end{proposition}

Subject to convenience, in what follows a generic HL flock shall be denoted by either
$[0, 1, \cdots, k]$ or $[1, \cdots, k]$. As in Cucker-Smale~\cite{CuckerSmaleA} or
Chung~\cite{chu94}, define the graph Laplacian matrix by
 \beginE \label{E2:lap}
  L=D- K, \qquad D=\mbox{diag}(d_0, \dots, d_{k}), \qquad d_i =\sum_{j} a_{ij}.
 \closeE
Similarly, define the two (non-orthogonally) complementing subspaces of $\bbR^{k+1}$:
\[
   \Delta=\mbox{span}\left\{ \begin{pmatrix} 1 \\ \vdots \\ 1 \end{pmatrix}_{k+1} \right\},
   \myand \bbR^{k}=\left\{ \begin{pmatrix} 0 \\ x_1 \\ \vdots \\ x_{k} \end{pmatrix} \mid
    x_i's \in \bbR \right\}.
\]
Then it is easy to see that
 \[ \Delta = \mbox{Ker}(L), \qquad \mbox{and $\bbR^k=\mbox{Range}(L)$ is $L$-invariant.}\]

Notice that the kernel assertion is directly guaranteed by the second condition of an HL
flock, without which the kernel could be larger.

From now one, as in Cucker and Smale~\cite{CuckerSmaleA,CuckerSmaleB}, we shall only
consider the restriction of the Laplacian on the reduced space $\bbR^k$. Then it becomes
nonsingular, and shall still be denoted by $L$ for convenience. {\em We also must point
out that when applied to actual flocking, the reduced Laplacian $L$ is applied to
$\bbR^{3k}$ (instead of $\bbR^k$) via the three spatial dimensions individually.}

\subsection{Definability of Compatible Inner Products}

The general framework of Cucker and Smale~\cite{CuckerSmaleA} relies upon the Fiedler
number of the Laplacian operator $L$, i.e., the smallest positive eigenvalue in the
reduced space. In particular, it assumes the existence of a fixed inner product
$\inner{\cdot}{\cdot}$ such that
 \beginE \label{E:positivity}
 \inner{Lv}{v} \ge \xi \inner{v}{v}, \qquad \mbox{for any}\; v \in \bbR^k.
 \closeE
 Then an {\em a priori \;} lower bound on $\xi=\xi(x)$ constitutes the core to the
 convergence results established by Cucker and Smale~\cite{CuckerSmaleA,CuckerSmaleB}.
 Below we show, however, that such inner products could fail to exist for non-symmetric
 systems like HL flocks.

\begin{theorem} \label{T:no-inner}
Consider the special HL $(k+1)$-flock $[0, 1, \dots, k]$ such that $\lead(i)=\{i-1\}$ for
$i>0$, and an instant when $a_{i,i-1} \equiv a$ for some fixed $a>0$ and any $i>0$. Then
the smallest eigenvalue is $\xi=a$, but there exists no inner product
$\inner{\cdot}{\cdot}$ in the reduced space $\bbR^k$ such that
 \[ \inner{Lv}{v} \ge a \inner{v}{v}, \qquad v \in \bbR^k. \]
\end{theorem}

\begin{proof}
It is easy to see that the (reduced) Laplacian $L$ is given by
 \[ L = L_a = \begin{bmatrix}
    a  & 0 & \dots & 0 & 0 \\
    -a & a & \dots & 0 & 0 \\
    \vdots & \vdots & \ddots & \vdots & \vdots \\
    0  & 0 & \dots & -a & a
        \end{bmatrix}_{k \times k}.
 \]
In particular, $L_a=a L_1$, and it suffices to prove the case when $a=1$. If such an
inner did exist, one would have
 \[ \inner{L_1 v}{v} \ge \inner{v}{v}, \qquad \mbox{or} \; \; \inner{Jv}{v} \ge 0,\]
where $J=L_1 - Id$. Notice that $Jv=(0, -z_1, \dots, - z_{k-1})^T$ for $v=(z_1, \dots,
z_k)^T$.

Let $e_i's$ denote the canonical basis of $\bbR^k$, and define
 \[ G=(g_{ij}) = (\inner{e_i}{e_j})_{k\times k} \]
to be the associated Grammian matrix of the inner product. Then $G$ must be positive
definite. For any $v=(z_1, \cdots, z_k)^T$, one has
 \[
 \inner{v}{Jv} = v^T G \cdot J v = (z_1, \dots, z_k) (g_{ij}) (0, -z_1, \dots,
 - z_{k-1})^T.
 \]
Consider a special vector in the form of $w=w_t =(0, \dots, 0, 1, t)^T \in \bbR^k$. Then
 \[
 \inner{w_t}{Jw_t} = (0, \dots, 0,1,t)(g_{ij})(0,\dots,0,1)^T = g_{k-1,k} + g_{k,k}t.
 \]
Notice that $g_{k,k}=\inner{e_k}{e_k} >0$. Then for any
 \[ t < - \frac{\left| g_{k-1,k} \right| }{g_{k,k}}, \]
one must have $\inner{w_t}{Jw_t} < 0$, which is contradictory. \byeproof
\end{proof}

Even when such compatible inner products do exist, for a general non-symmetric flock,
they often depend on the configuration of the flock, and are thus time-dependent. This
causes much inconvenience or a potential impasse for the Cucker-Smale approach
in~\cite{CuckerSmaleA,CuckerSmaleB}.  The efforts in the current work follow a different
approach by exploiting the specific structures of HL flocks.

\section{Discrete-Time Emergence}

Recall that in the continuous time, the Cucker-Smale flocking model is given by
 \begin{equation} \label{E:contime}
 \begin{cases}
 \dot x = v, \\
 \dot v = - L_x v, \qquad t > 0,
 \end{cases}
 \end{equation}
Where the reduced Laplacian $L=L_x$ is defined as in~(\ref{E2:lap}) and both $x$ and $v$
are considered in the reduced (quotient) space. For a $(k+1)$-flock, both of them belong
to $\bbR^{3k}$.

Fixing a discrete time step $h>0$. Define
 \[ x[n]=x(nh), \qquad v[n]=v(nh), \myand L_n = L_{x[n]}. \]
(Note: the parenthesis-bracket correspondence follows the convention in digital signal
processing~\cite{strngu}.) Then the continuous-time system~(\ref{E:contime}) is
discretized to
 \beginE \label{E3:disctime}
 \begin{cases}
 x[n+1]= x[n] + h v[n], \\
 v[n+1]= S[n] v[n], \qquad n=0, 1, \cdots
 \end{cases}
 \closeE
where $S[n] = S^h[n] = Id - hL_n$.

For an HL $(k+1)$-flock $[0, 1, \dots, k]$ , recall that the reduced Laplacian is given
by
 \beginE
  L_n = \begin{bmatrix}
    d_1[n]     & 0 & \dots & 0 & 0 \\
    -a_{21}[n] & d_2[n] & \dots & 0 & 0 \\
    \vdots     & \vdots & \ddots & \vdots & \vdots \\
    -a_{k1}[n] & -a_{k2}[n] & \dots & -a_{k,k-1}[n] & d_k[n]
        \end{bmatrix}_{k \times k}.
 \closeE
For $i >0$, since the leader set $\lead(i)\neq \kong$, we have
 \beginE \label{E:d}
 d_i[n] = \sum_{j=1}^k a_{ij}[n] = \sum_{j \in \lead[i]} a_{ij}[n] > 0.
 \closeE
Under the Cucker-Smale model, one has for any $j \in \lead(i)$,
 \beginE \label{E:a}
 a_{ij}[n] =
  \frac{H}{ \left( 1 + \left| \tilde x_j[n] - \tilde x_i[n] \right|^2/2 \right)^\b },
 \closeE
where $\tilde x_i$ denotes the original 3D position vector of agent $i$ (and the factor
$1/2$ is for convenience). In the reduced quotient space, one has $x_i = \tilde x_i -
\tilde x_0 \in \bbR^3$ since the original configuration vector $\tilde x \in
\bbR^{3(k+1)}$ and the reduced representation $x \in \bbR^{3k}$ are connected via:
 \[
 \tilde x =
 \begin{bmatrix}
 \tilde x_0 \\ \tilde x_1 \\ \vdots \\ \tilde x_k
 \end{bmatrix}
 =
 \begin{bmatrix}
 \tilde x_0 \\ \tilde x_0 \\ \vdots \\ \tilde x_0
 \end{bmatrix}
 +
 \begin{bmatrix}
 0 \\ \tilde x_1 - \tilde x_0 \\ \vdots \\ \tilde x_k - \tilde x_0
 \end{bmatrix}
=\begin{bmatrix}
 \tilde x_0 \\ \tilde x_0 \\ \vdots \\ \tilde x_0
 \end{bmatrix}
 + \begin{bmatrix}
  0 \\ x
  \end{bmatrix}.
\]
As a result, for any pair $i, j>0$,
 \[ |\tilde x_i - \tilde x_j|^2 = | x_i - x_j|^2 \le 2 (|x_i|^2 + |x_j|^2) \le 2  |x |^2.\]
In combination with~(\ref{E:d}) and~(\ref{E:a}), this implies that under the Cucker-Smale
connectivity,
 \beginE \label{E:d_lowerbound}
 d_i[n] \ge \frac {H} { (1 +  |x[n] |^2 )^\b } , \qquad i >0.
 \closeE

Assume, as in Cucker and Smale~\cite{CuckerSmaleA}, that under suitable initial
conditions (according to whether $\beta <, =, $ or $> \b_c=1/(2k)$), one has the uniform
bound on the reduced position vector:
 \beginE \label{E3:keybound}
   |x[n] |^2 \le B_h, \qquad \mbox{for} \; n=0, 1, \cdots,
 \closeE
 where $B_h$ is a constant bound depending only on $h$, the system parameters $H$ and $\b$,
  as well as the initial configuration.
 (The existence of $B_h$ is a crucial ingredient of the proof and will be further addressed
  immediately after this main line.)
 Then one has, for any $n\ge 0$, and $i>0$,
 \beginE \label{E:d_lowerbound}
  d_i [n] \ge d_\ast = \frac H {(1+B_h)^\b}.
 \closeE

\begin{proposition}[Uniform Elementwise Bound on S] \label{P:elementbound}
For $\displaystyle 0< h < \frac 1 {2 k H}$, $S_{ij}[n]\ge 0$ for any $i,j$, and
 \beginE \label{E:S_bound}
  \max_{i,j} S_{ij}[n] \le 1 - hd_\ast := \rho_h, \qquad n=0, 1, \cdots
 \closeE
\end{proposition}

\begin{proof}
By definition,
 \[
 S[n] = Id - h L_n =
  L_n = \begin{bmatrix}
    1-hd_1[n]     & 0 & \dots & 0 & 0 \\
    ha_{21}[n] & 1-hd_2[n]      & \dots & 0 & 0 \\
    \vdots     & \vdots         & \ddots & \vdots & \vdots \\
    ha_{k1}[n] & ha_{k2}[n]     & \dots & ha_{k,k-1}[n] & 1-hd_k[n]
        \end{bmatrix}_{k \times k}.
 \]
Under the condition on $h$, for the off-diagonals $i > j$, we have
 \[
 S_{ij}[n] = h a_{ij} \le h H < \frac 1 {2k} \le \frac 1 2.
 \]
For the diagonals, since $a_{ij} \le H$, we have $d_i \le (k-1)H$, and
 \[
 S_{ii}[n]= 1-h d_i \ge 1 - h (k-1)H > 1 - \frac 1 2= \frac 1 2.
 \]
Therefore,
 \[
 \max_{ij} S_{ij}[n] = \max_{i} S_{ii}[n]=1 - h \min_i d_i \le  1 -h d_\ast,
 \]
which completes the proof. \byeproof
\end{proof}

Next, our goal is to be able to control the growth rate of the matrix iteration:
 \[ S[n]S[n-1]\cdots S[0], \qquad \mbox{as $n \to \infty$.} \]
Normally, such asymptotic behavior is investigated via the so-called joint spectral
radius (e.g., Strang and Rota~\cite{RotaStrang60}, Daubechies and
Lagarias~\cite{DaubechiesMatrix92}, or Shen~\cite{Shen_LAA00,Shen_GeoMarkov00}):
 \[
  \lim_{n \to \infty} \left\|S[n]S[n-1]\cdots S[0]\right\|^{\frac 1 n},
 \]
which is often too complex to be feasible since the matrices evolve and generally do not
commute. The approach below resembles the Lebesgue Dominant Convergence Theorem in
Analysis~\cite{lielos}.

\begin{definition}[Domination]
A matrix $B=(b_{ij})$ is said to be dominated by another matrix $C=(c_{ij})$ of the same
size, if
 \[ |b_{ij}| \le c_{ij}, \qquad \mbox{for any}\; i, j. \]
If so, we write $B \prec C$.
\end{definition}

\begin{proposition} \label{P3:norm}
 If $B \prec C$, there exists some constant $\alpha$, such that
  \[  \|B \| \le \alpha \|C\|, \]
 where $\a$ only depends on the type of matrix norm adopted.
\end{proposition}

\begin{proof}
All norms in a finite-dimensional Banach space are equivalent. Therefore, it suffices to
establish the inequality under any special matrix norm. Consider the Fr\"obenius norm:
 \[
    \|B\|^2 = \mbox{trace}(BB^T) = \sum_{i,j} b_{ij}^2
    \le \sum_{i,j}c_{ij}^2 = \mbox{trace}(CC^T)=\|C\|^2,
 \]
with $\alpha=1$ (the superscript $T$ here denotes transpose). The general constant
$\alpha$ resurfaces when another norm is used instead.
 \byeproof
\end{proof}

\begin{proposition}
Suppose $B_i \prec C_i$, for $i=0, \dots, n$. Then
 \[ B_n B_{n-1} \cdots B_0 \prec C_n C_{n-1} \cdots C_0. \]
\end{proposition}

The proof is trivial. Next we define a ``complete" lower triangular matrix $T=(t_{ij})_{k
\times k}$ by
 \[
  t_{ij} = \begin{cases}
  1, & i \ge j; \\
  0, & \mbox{otherwise.}
  \end{cases}
 \]
Then the elementwise bound established in Proposition~\ref{P:elementbound} directly
implies the following.

\begin{corollary} Let $\rho_h = 1- hd_\ast$ as in Proposition~\ref{P:elementbound}. Then
\[  S[n] \prec \rho_h T, \myand
    S[n-1]\cdots S[0] \prec \rho_h^{n} T^n, \qquad n=0, 1, \dots
\]
\end{corollary}

\begin{lemma} \label{L:TnPolyGrow}
Let $T=(t_{ij})_{k \times k}$ be defined as above. Then $\|T^n\|= O(n^{k-1})$.
\end{lemma}

\begin{proof}
Denote by $J$ the $k$ by $k$ lower triangular matrix whose nonzero elements are all 1's
and {\em only} distributed right below the diagonal, e.g., the 3 by 3 case,
\[ J=
    \begin{bmatrix}
    0 & 0 & 0 \\
    1 & 0 & 0 \\
    0 & 1 & 0
    \end{bmatrix}.
\]
Then it is easy to see that
\[
 T = I + J + \cdots + J^{k-1}.
\]
Since $J^k=J^{k+1}=\cdots=0_{k\times k}$, one can also write
 \[
 T = \sum_{m=0}^\infty J^m.
 \]
More generally, for any $t$ with $|t| < 1$, one can define
 \[
 T(t) = \sum_{m=0}^\infty t^m J^m = ( I - tJ)^{-1}.
 \]
Then
 \[ T(t)^n = (I - tJ)^{-n} = \sum_{m=0}^\infty {{-n} \choose m} (-t)^m J^m
           = \sum_{m=0}^{k-1} { n+m-1 \choose m} t^m J^m.
 \]
Letting $t \to 1$, we have
 \[
 T^n = \lim_{t \to 1} T(t)^n =\sum_{m=0}^{k-1} { n+m-1 \choose m} J^m \prec O(n^{k-1}) T.
 \]
The proof is then complete via Proposition~\ref{P3:norm}. \byeproof
\end{proof}

Combining all the preceding results in this section, we have arrived at the following
conclusion.

\begin{theorem}  \label{T3:disc_convergence}
In the discrete-time Cucker-Smale model~(\ref{E3:disctime}) for an HL $(k+1)$-flock , for
any sufficiently small marching step $h$ (as in Proposition~\ref{P:elementbound} and
Cucker and Smale~\cite{CuckerSmaleA,CuckerSmaleB}), there exists some $\rho_h \in (0, 1)$
under the conditions similar to~\cite{CuckerSmaleA,CuckerSmaleB} based upon $\b <, =,$ or
$>\b_c=1/(2k)$, such that
 \[ S[n]\cdots S[0] \prec O(\rho_h^n n^{k-1}) T. \]
In particular, one has
 \[  |v[n] | \le O(\rho_h^n n^{k-1})  |v[0] |, \qquad n \to \infty. \]
The order constant in $O(\cdot)$ only depends on the size $k$ of the flock.
\end{theorem}

We point out that the polynomial growth rate $O(n^{k-1})$ (coming from $T^n$ in
Lemma~\ref{L:TnPolyGrow}) is characteristic of triangular HL flocks. A ``full" system
would make the approach here infeasible since
 \[
 \begin{bmatrix}
 1 & \cdots & 1 \\
 \vdots & \ddots & \vdots \\
 1 & \cdots & 1
 \end{bmatrix}_{k \times k}^n
 =\begin{pmatrix}
  1 \\ \vdots \\ 1
 \end{pmatrix}
  \begin{pmatrix}
   1 & \cdots & 1
  \end{pmatrix}
  \begin{pmatrix}
  1 \\ \vdots \\ 1
 \end{pmatrix} \cdots \cdots
 \begin{pmatrix}
   1 & \cdots & 1
  \end{pmatrix}
  =k^{n-1}\begin{bmatrix}
 1 & \cdots & 1 \\
 \vdots & \ddots & \vdots \\
 1 & \cdots & 1
 \end{bmatrix}.
 \]
The exponential growth rate $k^n$ would thus overpower $\rho_h^n$ and lead to an
exponential blowup.

Finally, we further address the important issue raised earlier in the proof concerning
the boundedness condition in (\ref{E3:keybound}): $|x[n]|^2 \le B_h$ for all $n$. The
existence of the convergence factor $\rho_h$ has crucially depended on such a bound
$B_h$. On the other hand, the very existence of $B_h$, as we intend to show now, depends
on $\rho_h$. This {\em entanglement} is characteristic of the nonlinear Cucker-Smale
flocking model (as well as in Vicsek et al.~\cite{Vicsek95} and Jadbabaie et
al.~\cite{JadbabaieLinMorse03}), and makes this type of models difficult to analyze. In
the rest of the section, we introduce the brilliant approach of Cucker and Smale in
unraveling such entanglement, which then genuinely completes the proof.

\begin{lemma}\label{L3:vtox}
For any given integer $k\ge 1$,
 \beginE \label{E3:vtox_lem}
 \sum_{m=0}^\infty t^m m^{k-1} \le (k-1)! (1- t)^{-k},  \qquad \forall \; t\in [0,1).
 \closeE
\end{lemma}

\begin{proof}
 Notice that the equality holds when $k=1$. Generally, for any $t \in [0, 1)$,
  \[
 \begin{split}
 (k-1)! ( 1- t)^{-k} & = (k-1)! \sum_{m=0}^\infty {-k \choose m} (-t)^m \\
    & = (k-1)! \sum_{m=0}^\infty {m+k-1 \choose k-1} t^m \\
    & = \sum_{m=0}^\infty (m+k-1)\cdots(m+1) t^m \\
    & \ge \sum_{m=0}^\infty m^{k-1} t^m,
 \end{split}
  \]
 which completes the proof. \byeproof
\end{proof}

We now apply the self-bounding technique developed by Cucker and Smale
in~\cite{CuckerSmaleA,CuckerSmaleB} to establish the bound $\big|x[n]\big|^2 \le B_h$
that is crucially needed in the proof of Theorem~\ref{T3:disc_convergence}. It also
explains the origin of the critical exponent $\b_c=1/(2k)$ and its role.

We thus return to the step in (\ref{E3:keybound}). This time, instead of assuming {\em a
priori} that $\big| x[n] \big|^2 \le B_h$ for {\em all \;} $n\ge 0$, we proceed as
follows. Fix any discrete time mark $N$, and define
 \beginE \label{E2:nstar}
 |x|_\ast = \max_{0 \le n \le N} |x|[n], \qquad N_\ast \in \mathrm{argmax}_{0 \le n \le N}
 |x|[n],
 \closeE
and similarly define
 \beginE \label{E3:N_dstar}
 d_\ast = \frac {H} { (1 + |x|_\ast^2)^\b }.
 \closeE
Thus $|x|_\ast$ could be considered as a ``localized" version of $B_h$, restricted in any
designated {\em finite} time segment $[0, N]$.

Then all the earlier analysis and results hold up to the bounding formula on $|v|[n]$ in
Theorem~\ref{T3:disc_convergence}, as long as one restricts $n$ within $[0, N]$. In
particular for $\rho_h = 1 - h d_\ast$,
 \[ |v|[n] \le A \rho_h^n n^{k-1}, \qquad n=0, \dots, N, \]
where the constant $A$ only depends on $k$ but on neither $n$ nor $N$.

Therefore, by the first equation of HL flocking in~(\ref{E3:disctime}), for any $n \in
[0, N]$,
 \[
 \begin{split}
  |x|[n] & \le |x|[0] + \sum_{m=0}^{n-1} |x[m+1] - x[m]|
          =   |x|[0] + h \sum_{m=0}^{n-1} |v[m]| \\
         & \le |x|[0] + A h \sum_{m=0}^{n-1} \rho_h^m m^{k-1}
          \le |x|[0] + A h \sum_{m=0}^\infty \rho_h^m m^{k-1} \\
         & \le |x|[0] + (k-1)! A h ( 1 - \rho_h )^{-k}.
 \end{split}
 \]
In particular, for $n=N_\ast$,
 \[
 |x|_\ast = |x|[N_\ast] \le |x|[0] + (k-1)! A h ( 1 - \rho_h)^{-k}.
 \]
Now that
 \[
 (1 - \rho_h)^{-k} = h^{-k} d_\ast^{-k} = (hH)^{-k} ( 1+ |x|_\ast^2)^{\b k},
 \]
one has the Cucker-Smale type of self-bounding inequality for the unknown $|x|_\ast$:
 \[
 |x|_\ast \le |x|[0] + (k-1)! Ah (hH)^{-k} ( 1 + |x|_\ast^2)^{\b k}.
 \]
Define $Z = (1 + |x|_\ast^2)^{1/2}$. Then
 \beginE \label{E3:csBound}
  Z \le 1 + |x|_\ast \le c + b Z^{2\b k},
 \closeE
with $\displaystyle c=1 + |x|[0]$ and $\displaystyle b = (k-1)! A h (hH)^{-k}$.

The rest of analysis then goes exactly as in Cucker and
Smale~\cite{CuckerSmaleA,CuckerSmaleB}. Define
 \[ F(z) = z - bz^s - c, \qquad \mbox{with} \;\; s = 2\b k, \myand z >0. \]
Then when $s < 1$, the nonlinear function $F(z)$ has a unique zero $z_\ast$ after which
$F$ stays positive. Since $F(Z) \le 0$, one thus must have $Z \le z_\ast$, or
 \[ |x|_\ast \le Z \le z_\ast. \]
Now that $z_\ast$ only depends on $c$ and $b$, which are independent of the pre-assigned
time mark $N$, we have obtained the uniform bound
 \[ |x|[N] \le |x|[N_\ast]=|x|_\ast \le z_\ast, \qquad \forall \; N=0, 1, \cdots. \]
Thus $B_h = z_\ast^2$ is the uniform bound needed in the proof of
Theorem~\ref{T3:disc_convergence}. This is the case when $\b \le \b_c =1/(2k)$.

The other two cases when $\b=\b_c$ and $\b>\b_c$ (corresponding to $s=1$ and $s>1$ for
$F(z)$) can be analyzed exactly in the same manner as in Cucker and
Smale~\cite{CuckerSmaleA,CuckerSmaleB}, and will be omitted herein. In particular, in
both cases, there will be {\em sufficient}-type of conditions on the initial
configurations in order for the bound $B_h$ to exist. In the third case $\b > \b_c$,
there will also be more stringent upper bound on the time marching size $h$. We refer the
reader to Cucker and Smale for the detailed analysis on $F(z)$ in these two cases. This
completes the proof of Theorem~\ref{T3:disc_convergence}.

In the next section, we investigate the emergent behavior of the continuous-time HL
flocking using quite different methods. There, the results hint that the unconditional
convergence range $\beta \in [0, 1/(2k))$ just established might still be extendable onto
$[0, 1/2)$, as in Cucker and Smale~\cite{CuckerSmaleA}. Thus the critical exponent
$\b_c=1/(2k)$ might be further improved if other alternative approaches are to be
investigated in the future.

\section{Continuous-Time Emergence}

Let $[1, \dots, k]$ be an HL $k$-flock in that order, connected via the Cucker-Smale
strength with parameters $\beta$ and $H$ as in~(\ref{E1:wchoice}). In this section, we
establish the emergence behavior for the entire flock when $\beta < 1/2$, via the methods
of induction and perturbation. The associated intuition is as follows. If the sub-flock
$[1, \dots, i-1]$ almost reaches convergence, it shall look like a rigid one-body to
agent $i$. Then $[1,\dots, i-1, i]$ is not far from a simpler two-agent flock. Our goal
is to develop rigorous mathematical analysis to quantify and support this point of
perspective. (In this section, we shall work with $[1,\dots, k]$ instead of $[0, 1,
\dots, k]$ due to the lack of advantage of introducing index 0.)

\subsection{The Property of Positivity}

The general properties to be established in this subsection are characteristic of the
Cucker-Smale flocking model. They could be useful for any future works on the model, on
top of their roles in the proof of the main results of this section.

Let $x_i, v_i \in \bbR^3$ denote the 3D position and velocity vectors of agent $i$.
Recall that the Cucker-Smale flocking model is given by
 \beginE \label{E:CScontT}
 \begin{cases}
 \dot x_i = v_i, \\
 \dot v_i = - (L_x v)_i = \sum_{j \in \lead(i)} a_{ij}(x) (v_j - v_i),
 \end{cases}
 \closeE
for $t>0$, $i=1,\dots,k$, and $x=(x_1, x_2, \dots, x_k) \in \bbR^{3k}$. The Cucker-Smale
connectivity strength is specified by
 \[
 a_{ij}(x)= \frac {H} { \left(1+ |x_j-x_i|^2\right)^\b}, \qquad j \in \lead(i).
 \]
 (As mentioned earlier in the Introduction, changing ``$=$" to ``$\ge$" does not
 affect the subsequent analysis as long as $a_{ij}(x)$'s are bounded and sufficiently
 smooth.)
Given a solution $(x(t), v(t))$ to the continuous Cucker-Smale model~(\ref{E:CScontT}),
we write for convenience
 \[ a_{ij}(t)=a_{ij}(x(t)), \myand L_t = L_{x(t)}.  \]
Let $\eta=(\eta_1, \eta_2, \cdots \eta_k)^T \in \bbR^k$ be $k$ scalars, and consider the
following system of ordinary differential equations:
 \beginE \label{E:eta_L}
 \dot \eta = - L_t \eta, \qquad t>0, \qquad \mbox{given} \; \eta^0 =\eta \big|_{t=0}.
 \closeE
Componentwise, we have
 \beginE \label{E:eta_A}
 \dot \eta_i = \sum_{j \in \lead(i)} a_{ij}(t) (\eta_j - \eta_i), \qquad i=1, \dots, k.
 \closeE

\begin{figure}[ht]
\centering{
 \epsfig{file=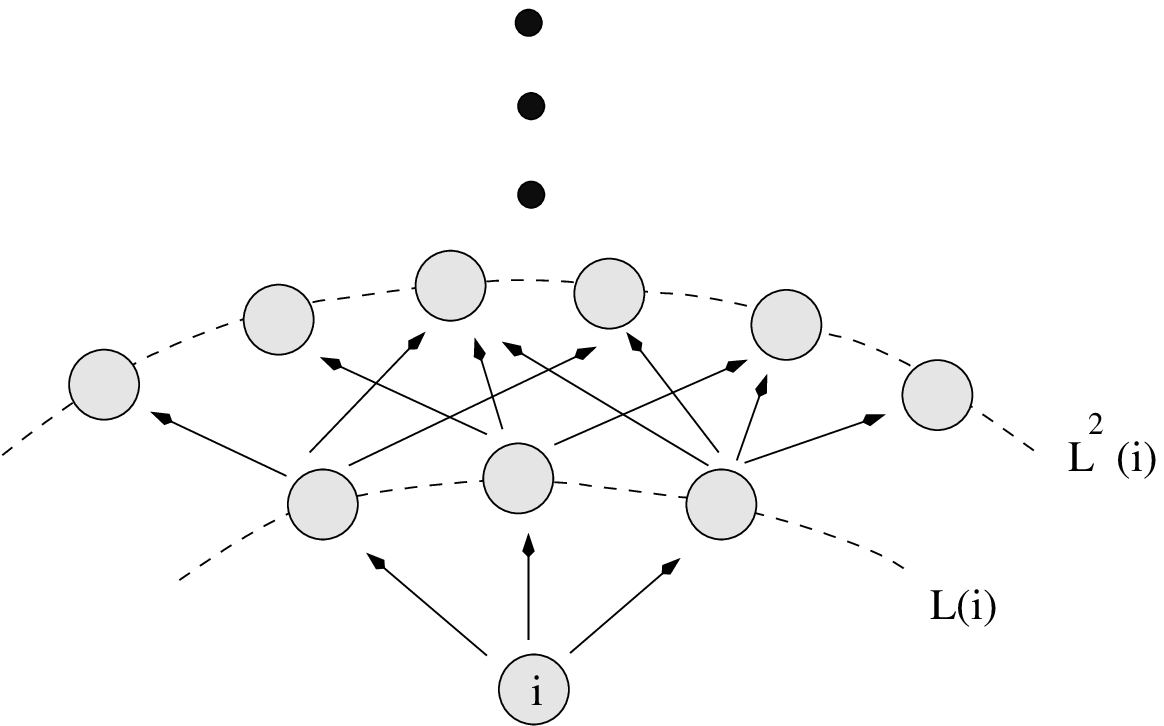, height=2in}
 \caption{The leaders of an agent $i$ at different levels:
    $\lead^0(i)=\{i\}, \lead(i), \lead^2(i), \cdots$.} \label{F2:leaderladder}
}
\end{figure}

\begin{theorem}[Positivity]
Suppose $\eta_i^0\ge 0$ for $i=1, \dots, k$. Then for all $t >0$ and $i$, $\eta_i(t)\ge
0$.
\end{theorem}

\begin{proof}
For any agent $i$ in the flock, define
 \beginE
 \begin{split}
 \lead^0(i) &=\{i\}, \\
 \lead^m(i) &=\lead(\lead^{m-1}(i)), \qquad \mbox{all $m$-th level leaders of $i$}, \quad
 \mbox{and} \\
 [\lead](i) &=\lead^0(i) \cup \lead^1(i) \cup \lead^2(i) \cdots, \qquad
 \mbox{all leaders of $i$, direct or indirect}.
 \end{split}
 \closeE
Then it is easy to see that the system (\ref{E:eta_A}) restricted on $[\lead](i)$ is
always self-contained, i.e., $(\eta_j \mid j \in [\lead](i))$ is not influenced by any
variables in $(\eta_j \mid j \notin [\lead](i))$ (but certainly not vice versa).

For convenience, we shall call the restriction of the system (\ref{E:eta_L}) or
(\ref{E:eta_A}) on the sub-flock $[\lead](i)$ {\em the $[\lead](i)$-system}. Then it
suffices to establish the theorem for each $[\lead](i)$ system.  In
Figure~\ref{F2:leaderladder}, we have sketched an example of the hierarchies of leaders
of a given agent $i$.

Suppose otherwise that the theorem were false on an $[\lead](i)$-system for some
particular agent $i$. There would exist some $\bar j \in [\lead](i)$, and $\bar t>0$ such
that $\eta_{\bar j}(\bar t)<0$. Define
 \[ t_\ast = \inf \{t > 0 \mid \mbox{there exists some}\;
             j \in [\lead](i), \;\mbox{such that}\; \eta_j(t)<0\}. \]
Then $0 \le t_\ast \le \bar t < \infty$, and we claim additionally the following.
 \begin{enumerate}[(i)]
 \item For any $j \in [\lead](i)$, $\eta_j(t_\ast) \ge 0$.
 \item There must exist some $\hat j \in [\lead](i)$, and a sequence of moments $(t_n)$
 such that $t_n > t_\ast$, $t_n \to t_\ast$ as $n\to \infty$, and $\eta_{\hat j}(t_n)<0$.
 \item There must exist some $j_\ast \in [\lead](i)$, such that $\eta_{j_\ast}(t_\ast)>
 0$.
 \end{enumerate}
(i) and (ii) result directly from the definition of $t_\ast$. Suppose otherwise (iii)
were false. Then in particular, for any $j \in [\lead](\hat j)$, one must have
$\eta_j(t_\ast) =0$ by (i). Consider the $[\lead](\hat j)$-system after $t_\ast$:
 \[
 \dot \eta_j = \sum_{l \in \lead(j)} a_{jl}(t)(\eta_l - \eta_j), \qquad j \in
 [\lead](\hat j), \;\; t > t_\ast.
 \]
Since this is a homogeneous system with zero initial conditions at $t=t_\ast$, by the
uniqueness theorem of ODEs (e.g.,~\cite{HirschSmaleODE}), the solution to the
$[\lead](\hat j)$-system must be identically zero: $\eta_j(t)\equiv 0$ for any $t>
t_\ast$ and $j \in [\lead](\hat j)$. Now that $\hat j \in [\lead](\hat j)$, one must have
$\eta_{\hat j}(t) \equiv 0$ for all $t > t_\ast$, which contradicts to Property (ii).
Thus (iii) holds.

Define
 \[
 \hat m = \min \{ m \ge 0 \mid \mbox{there exists some $j_\ast \in \lead^m(i)$, such
  that} \; \eta_{j_\ast}(t_\ast) > 0 \}.
 \]
Property (i) and (iii) imply that $0< \hat m < \infty$. Then by iteratively
differentiating the $\lead(\hat j)$-system, one can easily establish:
 \[
 \eta_{\hat j}(t_\ast)=\eta'_{\hat j}(t_\ast)=\cdots=\eta^{(\hat m-1)}_{\hat j}(t_\ast)=0,
 \qquad \eta^{(\hat m)}_{\hat j}(t_\ast) > 0,
 \]
which contradicts to Property (ii). Thus the theorem must hold and the proof is complete.
\byeproof
\end{proof}

The most important consequence is the following bounding capability.

\begin{theorem}[Boundedness of Velocities under Evolution]
 \label{T:boundV} The Cucker-Smale model~(\ref{E:CScontT}) has the following closedness
properties.
 \begin{enumerate}[(i)]
 \item Suppose $\Omega$ is a convex compact domain in $\bbR^3$, and for any agent $i$,
  initially $v_i(t=0) \in \Omega$. Then for any $t > 0$ and $i$, $v_i(t) \in
  \Omega$.
  \item In particular, let $D_0 = \max_i |v_i(t=0)|$. Then $|v_i(t)|\le D_0$ for all
  $t>0$ and $i$.
 \end{enumerate}
\end{theorem}

\begin{proof}
Since the closed ball $B_{D_0}(0)$ in $\bbR^3$ is convex and compact, (ii) is implied by
(i). It suffices to establish (i).

For any unit vector $n \in S^2$, and given vector $a \in \bbR^3$. We first claim that if
 \[ n \cdot (v_i - a) \big|_{t=0} \ge 0, \qquad \forall \; i, \]
then $n \cdot (v_i(t) - a) \ge 0$ remains valid for all $t>0$ and $i$. To proceed, define
$\eta_i =n\cdot(v_i-a)$.
 \[
 \begin{split}
 \dot \eta & = n \cdot \dot v_i \\
           & = n \cdot \left( \sum_{j \in \lead(i)} a_{ij}(t) (v_j-v_i) \right) \\
           & = n \cdot \left( \sum_{j \in \lead(i)}
                a_{ij}(t) \left[ (v_j-a)-(v_i-a)\right] \right) \\
           & = \sum_{j \in \lead(i)} a_{ij}(t) (\eta_j -\eta_i) \\
           & = - (L_t \eta)_i.
 \end{split}
 \]
Then by the preceding theorem, the claim is indeed valid: $\eta_i(t)\ge 0$ for all $t>0$
and $i$.

For any compact convex domain $\Omega$, let $p: S^2 \to \bbR^3$ be its support function,
so that for any unit direction $n \in S^2$, $a=p(n)$ has the property that $a \in \pd
\Omega$ and the closed flat half-space
 \[ \pi_{a, -n} =\{x \in \bbR^3 \mid (-n)\cdot (x -a) \ge 0\} \]
contains $\Omega$. When the domain is convex but not strictly convex, $p(n)$ could be a
set of points, which however does not influence the argument herein. Furthermore, we have
\[ \Omega = \bigcap_{n \in S^2} \pi_{p(n), -n}. \]
Since each half-space has just been shown invariant under the Cucker-Smale evolution, we
conclude that $\Omega$ must be invariant as well under the evolution, which completes the
proof. \byeproof
\end{proof}

\subsection{Perturbation and Induction}

We now first prepare a lemma. Together with the boundedness property just established
above, it facilitates the later analysis on the emergent behavior of HL flocks.

\begin{lemma} \label{L:perturb_system}
Suppose $x(t), v(t) \in \bbR^3$ (which could be considered as $x_2-x_1$ and $v_2-v_1$ for
a 2-flock), and satisfy the perturbed 2-flock system \underline{parametrized} by some
$T>0$:
 \beginE \label{E:perturbed}
 \begin{cases}
 \dot x = v (t) \\
 \dot v = - a_T(x, t) v(t) + \eps_T(t), \qquad  t\ge 0.
 \end{cases}
 \closeE
Assume in addition that the following conditions hold.
 \begin{enumerate}[(i)]
 \item $\displaystyle a_T(x, t) \ge \frac {H}{(1+|x|^2)^\b}$, with $\b < 1/2$.
 \item $\eps_T \in \bbR^3$, and
       \beginE \label{E:epsilon}
        \left| \eps_T(t) \right| \le a e^{-b(t+T)^\eta}, \qquad \mbox{for some
         $\eta \in (0, 1]$}.
       \closeE
 \item $|v(t)| \le D_0$ for all $t\ge 0$, and $|x_0| \le R_0 + D_0 T$.
 \end{enumerate}
Here $H, \b, a, b, \eta, D_0$, and $R_0$ are given constants independent of $T$. Let
$(x^T(t), v^T(t))$ denote the dependency on $T$. Then
 \beginE \label{E:perturbed_control}
 |v^T(T)| \le A e^{-B T^{(1-2\b)\wedge \eta^{-}}},
 \closeE
where $\eta^{-}=\eta-\delta$ for any small but positive $\delta$ when $\eta<1$, and
$\eta^-=1$ when $\eta=1$, and $A$ and $B$ are two constants only depending upon $H, \b,
a, b, \eta^-, D_0$, and $R_0$ (but not $T$).  The notation $a \wedge b$ represents
$\min(a, b)$.
\end{lemma}

\begin{remark} We first make two comments on the conditions.
\begin{enumerate}[(1)]
\item The all-time bound $|v(t)|\le D_0$ seems very stringent, but is now natural by
Theorem~\ref{T:boundV} in the preceding subsection.

\item As outlined in the beginning of the current section, the
lemma will be applied during the induction process going from the sub-flock
$[1,\dots,i-1]$ to $[1,\dots, i]$. To agent $i$, the  perturbation factor $\eps_T(t)$
comes from the exponentially small dispersion of the leading sub-flock $[1, \dots, i-1]$
from reaching exact emergence.
\end{enumerate}
\end{remark}
We now proceed to the proof of Lemma~\ref{L:perturb_system}.

\begin{proof}
From the equation for $v$, we have
 \[
 \begin{split}
 \inner{v}{\dot v} &= - a_T(x, t) \inner{v}{v} + \inner{v}{\eps_T(t)}, \qquad \mbox{or} \\
|v|\cdot|v|_t = \frac 1 2 \left(|v|^2\right)_t &= -a_T |v|^2 + \inner{v}{\eps_T(t)}.
 \end{split}
 \]
Assuming that $v$ is never identically zero on any non-empty open time interval (noticing
that the opposite scenario trivializes the lemma on any such intervals and the following
argument only needs a minor modification), one has
 \[
 \begin{split}
 |v|_t & \le - a_T |v| + |\eps_T| \\
       & \le - \frac H {(1+ |x|^2)^\b} |v| + a e^{-b (t+T)^\eta},
 \end{split}
 \]
by the conditions (i) and (ii). By $\dot x = v$ and (iii),
 \[ \begin{split}
 |x| & \le |x_0| + \int_0^t |v|(\tau) d\tau \\
     & \le R_0 + D_0 T + D_0 t = R_0 + D_0(t+T).
 \end{split} \]
As a result,
 \[
 |v|_t \le - \frac H { \left(1 + (R_0 + D_0(t+T))^2 \right)^\b} |v| + a e^{-b
 (t+T)^\eta}.
 \]
Then by the Gronwall-type integration,
 \[
 \begin{split}
 |v(t)| & \le |v_0|
    e^{-\int_0^t \frac H {\left(1 + (R_0 + D_0(\tau+T))^2 \right)^\b} d\tau}
           + a \int_0^t e^{-b(\tau+T)^\eta}
           \cdot e^{- \int_\tau^t \frac H {\left(1 + (R_0 + D_0(s+T))^2 \right)^\b} ds}
           d\tau \\
        & \le D_0 \cdot e^{- \frac {Ht} {\left(1 + (R_0 + D_0(t+T))^2 \right)^\b} } +
          \frac {a}{b\eta} T^{1-\eta} e^{-bT^\eta}.
 \end{split}
 \]
We denote $v(t)$ by $v^T(t)$ to indicate its dependency on $T$. Then
 \[ \begin{split}
 |v^T(T)| & \le D_0 \cdot e^{- \frac {H\cdot T} {\left(1 + (R_0 + 2D_0T)^2 \right)^\b} }
               + \frac {\tilde a(a, b, \eta^-)}{b \eta^-} e^{-b T^{\eta^-}} \\
         & \le D_0 e^{- \tilde H(H, R_0, D_0, \beta) T^{1-2\b}}
                + C(a, b, \eta^-) e^{-b T^{\eta^-}} \qquad \mbox{(when $T\ge 1$)} \\
         & \le A e^{-B T^{(1-2\b) \wedge \eta^-}},
 \end{split}
 \]
where the two constants $A$ and $B$ are independent of $T$. Also notice that when
$\eta=1$, the monomial factor $T^{1-\eta}=1$ and the lowering from $\eta$ to  $\eta^-$ is
unnecessary in the first line. Finally, since $|v^T(t)|\le D_0$ by the given conditions,
by suitably increasing $A$, the condition $T\ge 1$ in the last second line can actually
be removed. This completes the proof. \byeproof
\end{proof}

We are now ready to state and prove the main theorem.

\begin{theorem}[Convergence of an HL Flock] \label{T4:HLemergence}
Let $[1,2, \dots, k]$ be a Cucker-Smale flock under hierarchical leadership with $\beta <
1/2$. Then for some $B > 0$, which depends only on the initial configuration and all the
system parameters, one has
 \beginE\label{E:Vconvergence}
 \max_{1\le i,j \le k} |v_i(t) - v_j(t)| = O(e^{-Bt}), \qquad t > 0.
 \closeE
\end{theorem}

\begin{proof}
We prove the theorem by induction on the sub-flocks, from $[1,\dots, l-1]$ to $[1,\dots,
l]$.

First we show that the theorem holds for a 2-flock $[1,2]$. By definition, the leader set
$\lead(2)$ is nonempty and has to be $\lead(2)=\{1\}$, i.e., $a_{21}>0$. Let $x=x_2 -
x_1$, and $v=v_2 -v_1$. Then
 \[
 \begin{cases}
 \dot x =v \\
 \dot v= \dot v_2 - \dot v_1 = \dot v_2 = a_{21}(v_1 - v_2) = - a_{21} v.
 \end{cases}
 \]
Here $\displaystyle a_{21}=a_{21}(x)= \frac H {(1+|x|^2)^\b}$, with $\b < 1/2$. Then
Cucker and Smale's analysis in~\cite{CuckerSmaleA} still applies directly, and
$|v(t)|=O(e^{-Bt})$ for some $B>0$.

\begin{figure}[ht]
\centering{
 \epsfig{file=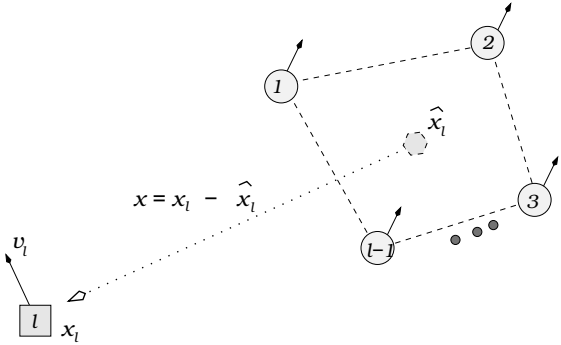, height=1.8in}
 \caption{The induction process from $[1,\dots,l-1]$ to $[1,\dots,l-1,l]$ reduces the
   $l$-flock system to a perturbed 2-flock system.}
}
\end{figure}

Assume now that the theorem holds for the sub-flock $[1, \dots, l-1]$, we intend to show
that it must be true for $[1,\dots, l-1, l]$ as well for $l>2$. As a result, the main
focus shall be the agent $l$.

By induction, there exists some $b>0$, such that
 \beginE \label{E:induct_l}
 \max_{i,j \in \{1,\dots,l-1\}} |v_i(t) - v_j(t)| = O(e^{-bt}), \qquad t \to \infty.
 \closeE
Define the average velocity (of the direct leaders of agent $l$) by
 \[
 \hat v_l(t) = \frac 1 d_l \sum_{i \in \lead(l)} v_i(t), \qquad
 \mbox{with} \; d_l = \# \lead[l].
 \]
Then for any $j \in \lead(l)$,
 \beginE \label{E:dispersionA}
 |v_j(t) - \hat v_l(t)| \le \frac 1 {d_l} \sum_{i \in \lead(l)} |v_j -v_i| = O(e^{-bt}),
 \closeE
by the induction assumption. Similarly, define
 \[
 \hat x_l(t) = \frac 1 {d_l} \sum_{i \in \lead(l)} x_i(t), \myand
 x(t)= x_l(t) - \hat x_l(t), \quad v(t) = v_l(t) - \hat v_l(t).
 \]
Then $\dot x = v$, and
 \beginE \label{E:v_induction}
 \begin{split}
 \dot v &= \dot v_l - \frac {d\hat v_l}{dt}
            =\sum_{j \in \lead(l)} a_{lj} \cdot (v_j - v_l) - \frac {d \hat v_l}{dt} \\
        &= \sum_{j \in \lead(l)} a_{lj} \cdot (\hat v_l - v_l)
            + \underbrace{
             \sum_{j \in \lead(l)} a_{lj} \cdot (v_j - \hat v_l)
            - \frac{d \hat v_l}{dt}
                }_{\eps(t)}.
 \end{split}
 \closeE
Since each $\dot v_i$ ($i \in \lead(l)$) is the linear combination of some $(v_j -v_i)$'s
with $j \in \lead(i) \subseteq \{1,\dots, l-1\}$, by~(\ref{E:induct_l}), one must have
 \[ \frac {d \hat v_l}{dt} = O(e^{-bt}). \]
Similarly, due to~(\ref{E:dispersionA}) and the boundedness of $a_{lj}$'s, one has
 \[
 \big| \sum_{j \in \lead(l)} a_{lj} \cdot (v_j - \hat v_l)  \big| = O(e^{-bt}).
 \]
In combination, we conclude that
 \beginE \label{E:eps_bounding}
 |\eps(t)| \le c e^{-bt}, \qquad t>0, \quad \mbox{for some}\; c>0.
 \closeE
On the other hand, define
 \beginE \label{E:def_a}
    a = \sum_{ j \in \lead(l)} a_{lj}
      = \sum_{j \in \lead(l)}
        \frac H {\left( 1 + |x_j - x_l|^2 \right)^\b }.
 \closeE
Then (\ref{E:v_induction}) simply becomes
 \beginE \label{E:a_eps}
 \dot v = - a v + \eps.
 \closeE

Define $\displaystyle g(s) = \frac H {(1+s)^\b}$ with $s \ge 0$. Then $g(s)$ is convex,
and
 \[
 \frac 1 {d_l} \sum_{j \in \lead(l)} g(s_j)
 \ge g\left(\frac 1 {d_l} \sum_{j \in \lead(l)} s_j \right).
 \]
As a result, when $s_j=|x_j - x_l|$,
 \beginE \label{E:convex_coalesce}
 \sum_{j \in \lead(l)} \frac H {( 1 + |x_j - x_l|^2)^\b} \ge
  d_l \frac {H} {\left( 1 + \frac 1 {d_l} \sum_{j \in \lead(l)} |x_j -x_l|^2 \right)^\b}.
 \closeE
By the least-square principle,
 \beginE \label{E:orthogonality}
 \frac 1 {d_l} \sum_{j \in \lead(l)} |x_l - x_j|^2 = |\overbrace{x_l - \hat x_l}^{x}|^2
 + \frac 1 {d_l}  \sum_{j \in \lead(l)} |x_j - \hat x_l|^2,
 \closeE
since $\hat x_l$ is the center or mean of $\{x_j \mid  j \in \lead(l)\}$. By the
induction assumption on the emergence of $[1,\dots,l-1]\supseteq \lead(l)$, there exists
some $M>0$, such that
 \beginE \label{E:bound_xvar}
 \frac 1 {d_l} \sum_{j \in \lead(l)} |x_j - \hat x_l|^2 \le M -1.
 \closeE
Combining Eqn.'s~(\ref{E:def_a}) through~(\ref{E:bound_xvar}), we have
 \beginE \label{E:lowerbound_a}
 a= a(x, t) \ge \frac {d_l H}{(M + |x|^2)^\b} \ge \frac {\tilde H}{(1 + |x|^2)^\b},
 \closeE
where the updated constant $\tilde H=\tilde H(H, d_l, M, \b)$. (Notice that the notation
$a(x,t)$ summarizes all the influence from $\{ x_j \mid j \in \lead(l)\}$ into the
$t$-variable.)

The combination of (\ref{E:eps_bounding}), (\ref{E:a_eps}), and (\ref{E:lowerbound_a})
leads to the reduced system:
 \beginE \label{E:reduced_2body}
 \begin{cases}
 \dot x = v \\
 \dot v = - a(x, t) v + \eps(t),
 \end{cases}
 \closeE
with $\displaystyle a(x,t) \ge \frac {\tilde H}{(1+|x|)^\b}, \myand
        |\eps(t)| \le c e^{-bt}$. In order to apply Lemma~\ref{L:perturb_system},
further define
 \beginE \label{E:DR}
 D_0 = 2 \max_{1 \le i \le k} |v_i (t=0)|, \myand
 R_0 = 2 \max_{1 \le i \le k} |x_i (t=0)|.
 \closeE
Then by Theorem~\ref{T:boundV}, we have
 \[
 |v_i(t)| \le \frac{D_0} 2, \qquad \forall \;  i \; \mbox{and}\; t>0.
 \]
Consequently,
 \beginE \label{E:vbyD}
 |v(t)| \le \frac 1 {d_l} \sum_{j \in \lead(l)} |v_j - v_l|
        \le \frac 1 {d_l} \sum_{j \in \lead(l)} D_0 = D_0.
 \closeE
Similarly, for any $T>0$,
 \[
 \begin{split}
 |x(T)-x(0)| & \le |x_l(T) - x_l(0)|
    + \frac 1 {d_l} \sum_{j \in \lead(l)} |x_j(T) - x_j(0)| \\
    & \le \frac {D_0} 2 T + \frac 1 d_l \sum_{j \in \lead(l)} \frac {D_0} 2 T = D_0 T.
 \end{split}
 \]
As a result,
 \beginE \label{E:xbyRD}
 \begin{split}
 |x(T)| & \le |x(0)| + D_0 T \\
        & \le |x_l(0)| + \frac 1 {d_l} \sum_{j \in \lead(l)}|x_j(0)| + D_0 T \\
        & \le \frac {R_0} 2 + \frac 1 {d_l} \sum_{j \in \lead(l)} \frac {R_0} 2 + D_0 T
        = R_0 + D_0 T.
 \end{split}
 \closeE

To conclude, for any $T>0$, if we define
  \[ x^T(t)= x(t+T), \quad v^T(t)=v(t+T), \quad a_T(x^T,t)=a(x^T, t+T), \;\;\mbox{and}
     \;\; \eps_T(t)=\eps(t+T),
  \]
then,
 \[
 \begin{cases}
 \dot x^T = v^T \\
 \dot v^T = - a_T(x^T, t) v^T + \eps_T(t), \qquad t> 0,
 \end{cases}
 \]
and all the three conditions in Lemma~\ref{L:perturb_system} are satisfied (with
$\eta=1$). Therefore, there must exist two positive constants $\tilde A$ and $\tilde B$,
such that for any $T>0$,
 \[
 |v(2T)|=|v^T(T)| \le \tilde A e^{-\tilde B T^{(1-2\b)\wedge 1}}
        = \tilde A e^{-\tilde B T^{1-2\b}}.
 \]
Since $T$ is arbitrary, we thus must have, after adjusting the constants,
 \[
 |v(t)| \le \hat A e^{- \hat B t^{1-2\b}}, \qquad t > 0, \qquad \mbox{for some constants
 $\hat A$ and $\hat B$}.
 \]

Moreover, since $\b < 1/2$ by assumption, one then must have
  \[ \int_0^\infty |v(t)| dt < \infty, \]
which in return implies that there exists some constant $M>0$, such that
  \[ |x(t)| \le M, \qquad t > 0. \]
Then by repeating the similar calculation in the proof of Lemma~\ref{L:perturb_system},
assisted with this new constant bound $|x|\le M$ instead of $|x|\le R_0 + D_0(t+T)$
there, one arrives at:
 \[
 |v(t)| \le A' e^{-B' t}, \qquad \mbox{(since $\eta=1$),}
 \]
for two positive constants $A'$ and $B'$ independent of $t$. Combined with the induction
base~(\ref{E:induct_l}), we thus conclude that the theorem must hold true for the
sub-flock $[1, \dots, l-1, l]$ with the exponent coefficient $B=B' \wedge b$. This
completes the proof. \byeproof
\end{proof}


\section{HL Flocking Under a Free-Will Leader}

In this section, partially inspired by the preceding perturbation methods, we investigate
a more realistic scenario when the ultimate leader agent 0 (in an HL flock $[0, 1, \dots,
k]$) can have a {\em free-will acceleration}, instead of merely flying in a constant
velocity.

The following phenomenon is not uncommon near lakes, grasslands, or any open spaces where
a flock of birds often visit. When the flock is initially approached by an unexpected
pedestrian or a predator from a corner on the outer rim, the bird which takes off first
(and alerts others subsequently) generally takes a curvy flying path before it reaches a
stable flying pattern with an almost constant velocity. Such a bird gains the full speed
fast, flies ahead of the entire flock, and serves as a virtual overall leader.

For an HL flock $[0, 1, \dots, k]$, in addition to the Cucker-Smale system
 \beginE \label{E5:csmodel}
 \begin{cases}
 \dot x_i = v_i(t) \\
 \dot v_i= \sum_{j \in \lead(i)} a_{ij}(x) (v_j(t) - v_i(t)), \qquad i>0,
 \end{cases}
 \closeE
we now also impose for the ultimate leader agent 0:
 \beginE \label{E5:freewillleader}
 \begin{cases}
 \dot x_0 = v_0 \\
 \dot v_0 = f(t), \qquad t > 0,
 \end{cases}
 \closeE
coupled with a given set of initial conditions. For convenience, we shall call $f(t)$ the
{\em free-will acceleration\,} of the leader. In combination, the new system is no longer
autonomous.

The main goal of this section is to establish the following theorem.

\begin{theorem} \label{T5:freewill_emerge}
Suppose an HL (k+1)-flock $[0, \dots, k]$ with a free-will leader satisfies
both~(\ref{E5:csmodel}) and (\ref{E5:freewillleader}), with the Cucker-Smale connectivity
strength of $\beta < 1/2$. In addition, assume that the leader's free-will acceleration
satisfies
 \[ |f(t)| = O((1+t)^{-\mu}), \qquad \mbox{with some exponent $\mu > k$.} \]
Then the flock still has the following emergent behavior:
 \[
 \max_{0 \le i, j \le k} |v_i - v_j|(t) = O\left( (1+t)^{-(\mu - k)} \right).
 \]
\end{theorem}

\begin{remark} We first make two comments regarding why one should expect to put some
               regularity conditions on the leader's behavior in order for a coherent
               pattern to emerge asymptotically.
 \begin{enumerate}[(1)]
 \item Intuitively, if the leader keeps changing its velocity substantially, it will be
 more difficult for the entire flock to follow and behave coherently. An extreme example
 is a flock with a {\em drunken} leader which flies in a Brownian random path. Then the
 entire flock cannot be expected to synchronize with the unpredictable motion of the leader
 instantaneously.

 \item In the theorem, the decaying constraint $\mu > k$ depends on the size $k$ of the
 flock. Thus qualitatively speaking, it requires the leader to exert less free will when
 the flock is larger, in order to lead a coherent flock asymptotically.
 Consider the special hierarchical leadership under a {\em linear chain of command}:
  \[ k \to k-1 \to \cdots \to 1 \to 0. \]
 The tail agent $k$ has to go through all the $k$ intermediate stages to sense any move
 that the leader is making. Thus intuitively, there will be a long time delay in between,
 and the leader has to be tempered enough to allow its distantly connected followers to
 respond coherently.
 \end{enumerate}
\end{remark}

We first prepare a lemma that is similar to Lemma~\ref{L:perturb_system}. Since the new
non-autonomous system does not necessarily have the positivity property, we take a
slightly different approach.

\begin{lemma} \label{L5:perturb}
Let $x, v, g \in \bbR^3$, and satisfy
 \[
 \begin{cases}
 \dot x = v(t) \\
 \dot v = -a(x, t) v(t) + g(t).
 \end{cases}
 \]
Suppose that
 \[
 \begin{split}
 a(x, t) &\ge \frac {H} {(1+|x|^2)^\b}, \qquad \mbox{for some $\b< 1/2$, and}, \\
 |g(t)| & = O \left( (1 + t)^{-\eta} \right), \qquad \mbox{with some constant $\eta > 1$}.
 \end{split}
 \]
Then, $|v(t)|=O((1+t)^{-(\eta -1)})$ with the order constant only depending on the
initial conditions $x(t=0), v(t=0)$, and $H$, $\b$, and $\eta$.
\end{lemma}

\begin{proof}
From the second equation, one has
 \[
 |v|\cdot|v|_t =\left(\frac{v^2}{2}\right)_t=\inner{v}{v_t}
           =- a \inner{v}{v}+\inner{v}{g}
           \le -a |v|^2 + |v|\cdot |g|
 \]
Assume that $v$ does not vanish identically on any non-empty open intervals for the same
reason as in the proof of Lemma~\ref{L:perturb_system}. Then one has
 \[ |v|_t \le - a |v| + |g|, \qquad t > 0. \]

Fix any time $T>0$, and define
 \beginE \label{E5:Tmax}
 |x|_\ast = \sup_{t \le T} |x|(t), \myand
 a_\ast= \inf_{t \le T} \frac H {(1 + |x|^2)^\b}= \frac H { (1+|x|^2_\ast)^\b. }
 \closeE
Then one has
 \beginE \label{E5:boundva}
  |v|_t \le - a_\ast |v| + |g|, \qquad t \in [0, T].
 \closeE
Since $a_\ast$ is constant, integration yields
 \[ |v|(t) \le |v|(0) e^{- a_\ast t} + \int_0^t |g|(\tau) e^{-a_\ast(t-\tau)} d\tau. \]
In particular, for any $t < T$,
 \[ |v|(t) \le |v|(0) + \int_0^t |g|(\tau) d\tau \le |v|(0)+ \int_0^\infty
 |g(\tau)|d\tau:=A_0.
 \]
(Since $\eta > 1$ by assumption, the integral of $|g|$ is finite.) Now that $A_0$ is
independent of the time mark $T$, we conclude that the last upper bound must hold for
{\em any} $t > 0$: $|v|(t)\le A_0, t > 0$. Therefore, from the first equation $\dot
x=v(t)$, one has
  \[ |x|(t) \le |x|(0) + \int_0^t |v|(\tau) d\tau\le B_0 + A_0 t, \quad t> 0, \]
where $B_0=|x|(0)$. In particular, for any time mark $T>0$, the quantities
in~(\ref{E5:Tmax}) are subject to:
 \[
 |x|_\ast \le B_0 + A_0 T, \myand a_\ast \ge \frac H {[1+ (B_0+A_0 T)^2]^\b}.
 \]
We then go back and integrate the inequality~(\ref{E5:boundva}) again, but from $T/2$ to
$T$ this time:
 \[
 \begin{split}
 |v|(T)
 & \le
 |v|(T/2) e^{-\frac{a_\ast T}{2}} + \int_{T/2}^T |g|(\tau) e^{-a_\ast (T-t)} dt \\
 & \le A_0 e^{ - \frac{HT/2}{[1+(B_0+A_0 T)^2]^\b} } + \int_{T/2}^\infty |g|(t) dt\\
 & \le A_0 e^{-\tilde H(A_0, B_0, \b) (1+T)^{1-2\b}} + \int_{T/2}^\infty
    O\left( (1+ t)^{-\mu} \right) dt \\
 &= A_0 e^{- \tilde H (1+T)^{1-2\b}} + O\left( (1+ T)^{-(\mu-1)} \right).
 \end{split}
 \]
Since $\b < 1/2$, we conclude that
\[
 |v|(T) = O\left( (1+ T)^{-(\mu-1)} \right),
\]
where the constant in $O(\cdot)$ is independent of $T$. Since $T$ is arbitrary, the lemma
is established. \byeproof
\end{proof}

We are now ready to prove Theorem~\ref{T5:freewill_emerge}. Details on some similar
calculations will be directed to the proof of Theorem~\ref{T4:HLemergence}.

\begin{proof}
It suffices to prove the following more general result:
 \beginE \label{E5:subflockConv}
 \max_{0 \le i, j \le l} |v_i - v_j|(t) = O\left( (1 + t)^{-(\mu-l)} \right),\qquad t>0,
 \closeE
for any sub-flock $[0, 1, \dots, l]$ and $l\ge 1$.

When $l=1$, define $x=x_1 - x_0$ and $v=v_1 - v_0$. Then $\dot x =v$, and
 \[ \dot v = \dot v_1 - \dot v_0 = a_{10} (v_0 - v_1)  - f=-a_{10}v -f. \]
By the definition of an HL flock, $\lead(1) \neq \varnothing$, and it has to be agent
$0$, implying that $a_{10}$ is subject to the Cucker-Smale formula. Then by the preceding
lemma (with $\eta =\mu$), one has
 \[ |v|(t) = O\left(  (1+t)^{-(\mu-1)}  \right), \]
and (\ref{E5:subflockConv}) holds.

Suppose now that (\ref{E5:subflockConv}) is true for the sub-flock $[0,1,\dots,l-1]$ with
$2 \le l \le k$, so that
 \beginE \label{E5:prel}
 \max_{0 \le i,j \le l-1} |v_i - v_j|(t) = O\left( (1+t)^{-(\mu -l +1)} \right).
 \closeE
As in the proof of Theorem~\ref{T4:HLemergence}, define the average features of the
direct leaders of agent $l$ by:
 \[
 \hat x_l = \frac 1 {d_l} \sum_{j \in \lead(l)} x_j, \myand
 \hat v_l = \frac 1 {d_l} \sum_{j \in \lead(l)} v_j, \qquad
 d_l = \# \lead(l),
 \]
and $x=x_l - \hat x_l$ and $v=v_l - \hat v_l$.

Then as in the proof of Theorem~\ref{T4:HLemergence}, one has $\dot x = v$ and
\[
 \begin{split}
 \dot v  &= - a(x, t) \cdot v + g_l(t), \qquad \mbox{with}\\
 g_l(t)  &= \sum_{j \in \lead(l)} a_{lj} \cdot (v_j - \hat v_l) - \frac {d\hat v_l }{dt}. \\
 a(x, t) &= \sum_{j \in \lead(l)} a_{lj}(x_l-x_j).
 \end{split}
\]
We first estimate $g_l$. Since $|a_{lj}| \le H$ and $\lead(l) \subseteq [0,1,\dots,
l-1]$, by the induction assumption~(\ref{E5:prel}), the first term in $g_l$ must be of
the order $O((1+t)^{-\eta})$ with $\eta=\mu - l +1$. For the remaining second term in
$g_l$, let $\displaystyle 1_{0\in \lead(l)}$ denote the logical variable which is 1 when
agent 0 belongs to $\lead(l)$, and 0 otherwise. Then
 \[
 \frac {d \hat v_l}{dt} = \frac 1 {d_l} \sum_{j \in \lead(l)} \dot v_j
  = 1_{0 \in \lead(l)} \cdot \frac 1 {d_l} \dot v_0
   + \frac 1 {d_l} \sum_{j \in \lead(l) \setminus \{0\}} \dot v_j.
 \]
Now that $\dot v_0 =f(t)=O((1+t)^{-\mu})$, and each $\dot v_j$ with $j \in \lead(l)
\setminus \{0\}$ is some linear combination of $(v_s-v_j)$ with $s$'s in $\lead(j)
\subseteq [0, 1, \dots, l-1]$. Thus by the induction assumption~(\ref{E5:prel}), one must
have $\frac {d\hat v_l}{dt} = O((1+t)^{-\eta})$ with $\eta= \mu - l+1$.

We now estimate $a(x,t)$. Since $\mu > k$ by the given condition, we have $\mu - l + 1>
k-l +1 =1$. As a result, by the induction assumption on the sub-flock $[0, 1, \dots,
l-1]$, for any $i, j \le l-1$,
 \[
 \begin{split}
 |x_i - x_j|(t) & \le |x_i - x_j|(0) + \int_0^t |v_i - v_j|(\tau) d\tau \\
                & \le |x_i - x_j|(0) +
                \int_0^\infty O\left( (1+\tau)^{-(\mu-l+1)}\right) d\tau < \infty,
                \qquad \forall \; t>0.
 \end{split}
 \]
Therefore the boundedness property in~(\ref{E:bound_xvar}) still holds, and the same
calculation in the proof of Theorem~\ref{T4:HLemergence} leads to
 \[
 a(x, t) \ge \frac {\tilde H} {(1 + |x|^2 )^\b},
 \]
for some constant $\tilde H=\tilde H(H, d_l, \b, f, \mbox{initial conditions of}\;
[0,\dots,l-1])$.

Combining the estimations on $g_l$ and $a$, one sees that $x(t)$ and $v(t)$ satisfy a
perturbed system as in Lemma~\ref{L5:perturb} with $\eta=\mu - l +1$. Therefore, by
Lemma~\ref{L5:perturb},
 \[
 |v_l-\hat v_l|(t)= |v|(t) = O\left( (1+ t)^{-(\eta-1)}\right)=O\left( (1+t)^{-(\mu-l)} \right).
 \]
Now that by the induction assumption, for any $j \le l-1$, one must have
 \[
 |v_j - \hat v_l|(t) =O\left( (1+t)^{-(\mu-l+1)} \right), \;\; \mbox{since} \;
 |v_j - v_i|=O\left( (1+t)^{-(\mu-l+1)} \right), \;\; \forall \; i \in \lead(l).
 \]
Therefore, for any $j \le l-1$,
 \[
 |v_l - v_j| \le |v_l - \hat v_l| + |v_j - \hat v_l| = O\left( (1+t)^{-(\mu-l)}
 \right).
 \]
This completes the proof of~(\ref{E5:prel}), and thus the entire theorem. \byeproof
\end{proof}

\begin{corollary}
Under the same statements as in the preceding theorem, suppose $\mu > k+1$, then there
exists a constant configuration $\displaystyle (d_{ij})_{0\le i,j\le k}$ with $d_{ij} \in
\bbR^3$, such that
 \[
  \lim_{t \to \infty} \left(x_i(t) - x_j(t)\right) = d_{i,j}, \qquad 0 \le i, j \le k,
 \]
and the convergence rate is $O\left( (1+t)^{ -(\mu - k -1) } \right)$.
\end{corollary}

\section{Conclusion}

In this paper, we have investigated the emergent behavior of Cucker-Smale flocking under
the structure of {\em hierarchical leadership}\; (HL). The convergence rates are
established for the general cases of both discrete-time and continuous-time HL flocking,
as well as for HL flocking under an overall leader with free-will accelerations.

In all these cases, the consistent convergence towards some asymptotically coherent
patterns may reveal the advantages and necessities of having leaders and leadership in a
complex (biological, technological, economic, or social) system with sufficient
intelligence and memory.

Our future work shall focus more on extending the results herein onto other flocking
systems or leadership structures, a few of which have been mentioned in the Introduction.

\section*{Acknowledgments}

The author thanks both  Professors Steve Smale and Felipe Cucker for their generosity in
sharing the ideas on their developing frameworks. The work is impossible without the
patient daily guidance, encouragement, and numerous suggestions from Prof. Smale. The
author is profoundly grateful for the tender care from Prof. Smale and Prof. David
McAllester, as well as for the generous visiting support from the Toyota Technological
Institute (TTI-C) on the campus of the University of Chicago during the fall semester of
2006. The author also thanks his Ph.D. advisor Prof. Gil Strang for the timely directing
to the works of Prof. Iain Couzin on effective leadership.



\end{document}